# Test of relativistic gravity for propulsion at the Large Hadron Collider

Franklin Felber

*Starmark, Inc.*
*P. O. Box 270710*
*San Diego, CA 92198*
*858-676-0055; felber@san.rr.com*

**Abstract.** A design is presented of a laboratory experiment that could test the suitability of relativistic gravity for propulsion of spacecraft to relativistic speeds. An exact time-dependent solution of Einstein's gravitational field equation confirms that even the weak field of a mass moving at relativistic speeds could serve as a driver to accelerate a much lighter payload from rest to a good fraction of the speed of light. The time-dependent field of ultrarelativistic particles in a collider ring is calculated. An experiment is proposed as the first test of the predictions of general relativity in the ultrarelativistic limit by measuring the repulsive gravitational field of bunches of protons in the Large Hadron Collider (LHC). The estimated 'antigravity beam' signal strength at a resonant detector of each proton bunch is 3 nm/s$^2$ for 2 ns during each revolution of the LHC. This experiment can be performed off-line, without interfering with the normal operations of the LHC.

**Keywords:** Gravitational propulsion, relativistic gravity, 'antigravity', exact field solutions, experimental tests of gravity, Large Hadron Collider

**PACS:** 04.80.-y, 04.80.Cc, 04.20.-q, 04.20.Jb

## INTRODUCTION

This paper presents a design of a laboratory experiment that could test the suitability of relativistic gravity for propulsion of spacecraft to relativistic speeds. Within the weak-field approximation of general relativity, exact solutions have been derived for the gravitational field of a mass moving with arbitrary velocity and acceleration (Felber, 2005a). The solutions indicated that a mass having a constant velocity greater than $3^{-1/2}$ times the speed of light $c$ gravitationally repels other masses at rest within a narrow cone. At high Lorentz factors ($\gamma \gg 1$), the force of repulsion in the forward direction is about $-8\gamma^5$ times the Newtonian force. This strong dependence of the repulsive force on $\gamma$ offers opportunities for laboratory tests of gravity at extreme velocities. This paper outlines such an experiment, measuring with a resonant detector the force of gravity produced by protons in a collider ring.

An exact time-dependent solution of Einstein's gravitational field equation (Felber, 2008 and 2009) confirms the weak-field result that even the weak field of a mass moving faster than $3^{-1/2}c$ is repulsive in the forward and backward directions. This exact 'antigravity-field' solution was calculated from an exact metric first derived, but not analyzed, by (Hartle, Thorne and Price, 1986). The exact results confirm that a large mass moving faster than $3^{-1/2}c$ could serve as a driver to accelerate a much smaller payload from rest to a good fraction of the speed of light.

The exact results are consistent with the repulsion of relativistic particles by a static Schwarzschild field, discovered by (Hilbert, 1917 and 1924). That a particle with a radial speed exceeding $3^{-1/2}c$ is repelled in a *static* Schwarzschild field was first correctly noted by (Hilbert, 1917), as shown in Figure 1. A recent historical note (Loinger and Marsico, 2009) provides an informal summary translation and a discussion of Hilbert's remarkable but little-known paper. Subsequent papers have addressed (Carmeli, 1972) and reviewed (McGruder, 1982) the critical speed for radial motion in a Schwarzschild field. The same critical speed of repulsion was found for radial motion along the rotation axis of a spinning stationary source in (Mashhoon, 2005).

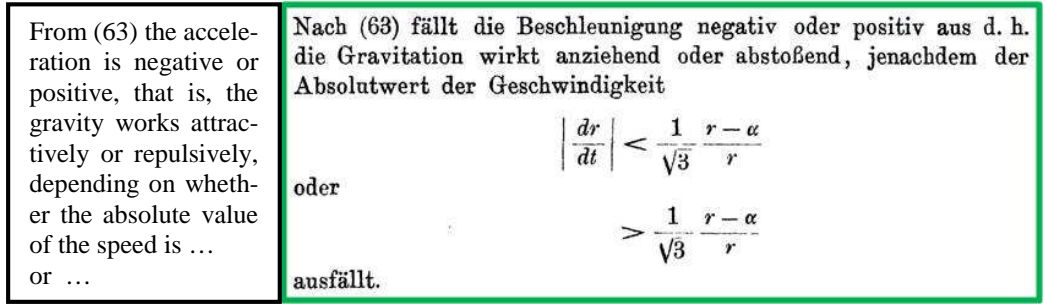

**Figure 1.** Repulsion in Schwarzschild field above $3^{-1/2}c$ from (Hilbert, 1917).

The discovery in (Felber, 2005a) of a repulsive weak gravitational field produced by relativistic masses was used in (Felber, 2005b and 2006a) to calculate the exact relativistic motion of a particle in the strong gravitational field of a mass moving with constant relativistic velocity, but without an explicit calculation of the strong dynamic gravitational field that produced the motion. (Felber, 2005b, 2006a, 2006b and 2006c) then showed how even a weak repulsive field of a suitable driver mass at relativistic speeds could quickly propel a heavy payload from rest to a speed significantly faster than the driver and close to the speed of light, and do so with manageable stresses on the payload.

The use of 'antigravity' fields of large relativistic masses for spacecraft propulsion was first proposed in (Felber, 2005b and 2006a). In these papers, it was shown that the propulsion of spacecraft to relativistic speeds could be calculated simply by a Lorentz transformation of the known unbound orbits of a spacecraft in a Schwarzschild field to a reference frame in which the large driver mass was moving at a constant relativistic velocity. In (Felber, 2006b and 2006c), it was shown by the same methods that a relativistic driver could deliver a specific impulse to a spacecraft much greater than the specific momentum of the driver, a condition denoted 'relativistic hyperdrive'. The advantages of 'antigravity' propulsion for near-light-speed travel, if such a driver could be found and accessed in space, are: (i) The daunting energy needs would be available naturally; and (ii) the stresses on the payload from acceleration would be limited to manageable tidal forces in 'free fall' motion along a geodesic.

This paper presents a design of an experiment to test these predictions of general relativity in the ultrarelativistic limit. The experiment, a type first proposed by (Braginsky, Caves and Thorne, 1977) for a Tevatron-scale storage ring, measures the gravitational field of bunches of protons in a storage ring. Our estimates suggest that each proton bunch in the Large Hadron Collider (LHC) beam would produce an 'antigravity beam' with a signal strength of 3 nm/s$^2$ and a duration of 2 ns at a detector. With a suitable high-$Q$ resonant monocrystal detector and a typical proton circulation time of 10 hours, the sound pressure level of the 31.6-MHz 'antigravity beam' at the LHC could be resonantly amplified to exceed 160 dB *re* 1 µPa. This type of experimental test is particularly attractive because it can be performed off-line, without interfering with the normal operations of the LHC.

This type of test is also attractive because it can be used to test general relativity in the ultrarelativistic regime, as well as to assess the potential of relativistic 'antigravity' for propulsion of payloads in the distant future. Tests of general relativity can be performed in laboratory experiments with much less difficulty and cost, and in less time, than comparable space experiments. For example, the National Aeronautics and Space Administration (NASA) began funding Gravity Probe B in 1964, intending to test the geodetic effect and inertial frame dragging at nonrelativistic speeds (http://www.nasa.gov/mission_pages/gpb/index.html). It was launched 40 years later. By the time of launch, the total cost was $750 million. Following multiple delays in data analysis over several years (Kestenbaum, 2007; NASA, 2008), the total cost may by now have exceeded $1 billion. Tests of relativistic gravity at existing accelerator facilities can be performed for less than 1% of that cost.

## WEAK 'ANTIGRAVITY' FIELDS

The design of experiments to test relativistic gravity for propulsion depends on gravitational fields only in the weak-field approximation of general relativity. This section reviews the calculation (Felber, 2005a) in the weak-field approximation of general relativity of the retarded gravitational field in conventional 3-vector notation of a particle mass having any arbitrary motion. Although the newly derived *exact* time-dependent gravitational-field solution of Einstein's equation (Felber, 2008 and 2009) is not needed for designing laboratory tests of relativistic gravity, it is reviewed nevertheless in the next section as a way of confirming the weak-field results of this section, which some might otherwise attribute to spurious coordinate effects.

Earlier calculations (Misner, Thorne and Wheeler, 1973; Braginsky, Caves and Thorne, 1977; Harris, 1991; Ohanian and Ruffini, 1994; Ciufolini and Wheeler, 1995) of the gravitational fields of arbitrarily moving particles were done only to first order in $\beta = u/c$, the source velocity normalized to the speed of light. But since the repulsive-force terms are second-order and higher in $\beta$, this 'antigravity' had not previously been found.

The relativistically exact bound and unbound orbits of test particles in the strong *static* field of a mass at rest have been thoroughly characterized in (Misner, Thorne and Wheeler, 1973; Chandrasekhar, 1983; Ohanian and Ruffini, 1994), for example. Even in a weak static field, earlier calculations of fields (Misner, Thorne and Wheeler, 1973; Braginsky, Caves and Thorne, 1977; Harris, 1991; Ohanian and Ruffini, 1994; Ciufolini and Wheeler, 1995) only solved the geodesic equation for a *nonrelativistic* test particle in the *slow-velocity* limit of source motion. In this slow-velocity limit, the field at a *moving* test particle has terms that look like the Lorentz field of electromagnetism, called the 'gravimagnetic' or 'gravitomagnetic' field (Harris, 1991; Ohanian and Ruffini, 1994; Ciufolini and Wheeler, 1995). (Harris, 1991) derived the *nonrelativistic* equations of motion of a moving test particle in a dynamic field, but only the dynamic field of a *slow-velocity* source. (Mashhoon, 2008) calculated the dynamic gravitomagnetic field of a slowly spinning source having slowly varying angular momentum, and more generally showed explicitly that general relativity contains induction effects at slow source velocities (Bini *et al*., 2008).

An exact solution of the field of a relativistic mass is the Kerr solution (Kerr, 1963; Boyer and Lindquist, 1967; Misner, Thorne and Wheeler, 1973; Chandrasekhar, 1983; Ohanian and Ruffini, 1994), which is the exact *stationary* solution for a rotationally symmetric spinning mass. Although time-independent, the Kerr field exhibits an inertial-frame-dragging effect (Misner, Thorne and Wheeler, 1973) similar to that contributing to gravitational repulsion at relativistic velocities. In the stationary Kerr gravitational field, the relativistic unbound orbits of test particles have been approximated in (Barrabès and Hogan, 2004).

This section summarizes the calculation in (Felber, 2005a) of a relativistically exact nonstationary field within the weak-field approximation of general relativity. A Liénard-Wiechert "retarded solution" approach (Kopeikin and Schäfer, 1999) is used to solve the linearized field equations from the retarded Liénard-Wiechert tensor potential of a relativistic particle. The solution is used to calculate the weak field acting on a test particle at rest of a mass moving with arbitrary relativistic motion. Since this solution is the first ever used to analyze the field of a translating mass beyond first order in $\beta$, it is the first to reveal that a mass having a constant velocity greater than $3^{-1/2}c$ gravitationally repels other masses at rest within a cone, as seen by a distant inertial observer. The (Aichelburg and Sexl, 1971) solution and other boosted solutions, such as in (Barrabès and Hogan, 2004), apply only in the proper frame of the accelerating particle, in which no repulsion appears, and not in the laboratory frame.

A Liénard-Wiechert "retarded solution" approach was used by (Felber, 2005a) to solve for the *dynamic* field of a relativistic particle in arbitrary motion. By using the retarded Green's function to solve the linearized field equations in the weak-field approximation, (Kopeikin and Schäfer, 1999) calculated the exact "retarded Liénard-Wiechert tensor potential" of a relativistic particle of rest mass $m$,

$$h_{\mu\nu}(\mathbf{r_0}, t) = \frac{-4Gm}{c^4} \int \frac{S_{\mu\nu}(t')}{\gamma(t')r(t')} \delta\left(t' + \frac{r(t')}{c} - t\right) dt' = \frac{-4Gm}{c^4} \left\{\frac{S_{\mu\nu}}{\gamma \kappa r}\right\}_{\text{ret}}. \tag{1}$$

In equation (1), the metric tensor was linearized as $g_{\mu\nu} \approx \eta_{\mu\nu} + h_{\mu\nu}$, where $\eta_{\mu\nu}$ is the Minkowski metric tensor; $S_{\mu\nu} \equiv u_\mu u_\nu - c^2 \eta_{\mu\nu}/2$ is a source tensor, with pressure and internal energy neglected; $u^\mu = \gamma(c, \mathbf{u})$ is the 4-vector velocity of the source; $\mathbf{u}$ is the 3-vector velocity; and $\gamma = (1-\beta^2)^{-1/2}$ is the Lorentz (relativistic) factor. As shown in Figure 2, $\mathbf{r}' = \mathbf{r_0} - \mathbf{s}'$ is the displacement vector from the source position $\mathbf{s}'(t')$ to the detector or test particle at $\mathbf{r_0}(t)$; $\hat{\mathbf{n}}' = \mathbf{r}'/r'$ is a unit vector; the delta function provides the retarded behavior required by causality; the factor $\kappa \equiv 1 - \hat{\mathbf{n}} \cdot \mathbf{u}/c$ is the derivative with respect to $t'$ of the argument of the delta function, $t' + [r'(t')/c] - t$; and the quantity in brackets $\{\ \}_{\text{ret}}$ and all primed quantities are to be evaluated at the retarded time $t' = t - r'/c$.

Equation (1) is the starting point in (Felber, 2005a) for the exact calculation of the gravitational field of a relativistic particle from the tensor potential. In the weak-field approximation, the retarded tensor potential of equation (1) is exact, even for relativistic velocities of the source. And since the tensor potential is linear, the field to be derived from it in this paper is easily generalized to ensembles of particles and to continuous source distributions. A treatment of post-linear corrections to the Liénard-Wiechert potentials in (Kopeikin and Schäfer, 1999) is given in (Kovács and Thorne, 1978) and references therein.

From equation (1), (Felber, 2005a) calculated the relativistically exact, but weak, retarded gravitational field of a moving source on a test particle instantaneously at rest at $(\mathbf{r_0},t)$ as

$$\mathbf{g}(\mathbf{r_0},t) = -Gm\left\{\frac{(\alpha/\gamma^2)\hat{\mathbf{n}} + [(2\gamma+1/\gamma)\kappa - 4/\gamma]\boldsymbol{\beta}}{\kappa^3 r^2} + \frac{(\hat{\mathbf{n}}\cdot\dot{\boldsymbol{\beta}})(\alpha\hat{\mathbf{n}} - 4\gamma\boldsymbol{\beta}) + \kappa(\dot{\alpha}\hat{\mathbf{n}} - 4\dot{\gamma}\boldsymbol{\beta} - 4\gamma\dot{\boldsymbol{\beta}})}{c\kappa^3 r}\right\}_{\text{ret}}, \quad (2)$$

where $\alpha \equiv 2\gamma - 1/\gamma$ and an overdot denotes differentiation with respect to $t'$. Just as does the retarded electric field of a point charge (Jackson, 1962 and 1975), the weak retarded gravitational field divides itself naturally into a 'velocity field', which is independent of acceleration and which varies as $r^{-2}$, and an 'acceleration field', which depends linearly on $\dot{\beta}$ and which varies as $r^{-1}$. The radial component of the electric 'velocity field' never changes sign. The radial component of the gravitational 'velocity field', on the other hand, can change sign at sufficiently high source velocity and repel masses within a narrow cone.

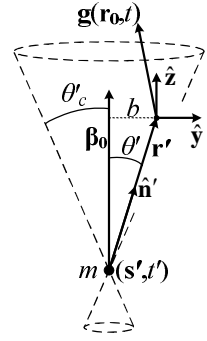

**Figure 2.** Configuration of source mass $m$ acting on test particle at rest at $(\mathbf{r_0},t)$.

Equation (2) is the gravitational field that would be observed by a distant inertial observer to act on a test particle at rest at $(\mathbf{r_0},t)$. If the test particle moves with velocity $\mathbf{v}$, then the gravitational field measured by the moving test particle has additional 'gravimagnetic' terms. The following calculations apply in the rest frame of the test particle and use the impulse approximation, which has the test particle remaining at rest during the time the gravitational field of the moving particle acts upon it. (Any difference in fields caused by motion induced in the test particle by the source is of the same order as terms that have already been neglected in the weak-field approximation.)

A particle of mass $m$ moving with constant velocity $\boldsymbol{\beta_0}c$ and $\gamma_0 = (1-\beta_0^2)^{-1/2}$ at impact parameter $b$ with respect to a test particle at rest at $(\mathbf{r_0},t)$ and retarded polar angle $\theta'$, as shown in Figure 2, produces a gravitational field at the test particle with a radial component,

$$g_r = \mathbf{g}\cdot\hat{\mathbf{n}}' = -Gm\left\{\frac{1-\beta_0^4 - [1-3\beta_0^2 + (3-\beta_0^2)\beta_0\cos\theta]\beta_0\cos\theta}{(1-\beta_0^2)^{1/2}(1-\beta_0\cos\theta)^3 r^2}\right\}_{\text{ret}}. \quad (3)$$

Figure 3 shows as a function of $\beta_0$ the retarded critical polar angle,

$$\theta_c' = \cos^{-1}\left[\frac{3\beta_0^2 - 1 + (13 - 10\beta_0^2 - 3\beta_0^4 + 4\beta_0^6)^{1/2}}{2\beta_0(3-\beta_0^2)}\right], \quad (4)$$

inside which $g_r$ repels stationary masses, instead of attracting them. Analysis of equation (4) shows that $\beta_0 > 3^{-1/2}$ and either $\theta' < 23.844$ deg or $\theta' > 180 - 23.844$ deg are both necessary conditions for gravitational repulsion of stationary masses by a particle having constant velocity. For $\gamma_0 \gg 1$, the 'antigravity beam' divergence narrows to $\theta_c' \approx 1/\gamma_0$. For $\gamma_0 \gg 1$, we also found in (Felber, 2005a) that $g \approx -8\gamma_0^5 g_N$ in the forward direction ($\theta'=0$), where $g_N \equiv -Gm/r'^2$ is the Newtonian field. For $\gamma_0 \gg 1$, the repulsive field in the backward direction ($\theta'=\pi$), $g \approx -(\gamma_0/2)g_N$, is weaker by a factor $(2\gamma_0)^{-4} \ll 1$.

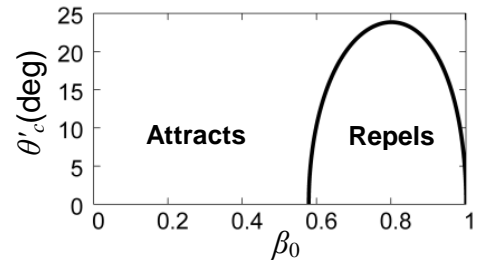

**Figure 3.** Retarded half-angle of 'antigravity beam' vs. $\beta_0$.

If a mass *m* approaching or receding from a stationary test particle at a speed greater than $3^{-1/2}c$ repels the test particle, then a stationary mass *M* should also repel a test particle that approaches or recedes from it at a speed greater than $3^{-1/2}c$. In the Schwarzschild field of a stationary mass *M*, the exact equation of motion of a test particle having *purely radial motion* (zero impact parameter) is

$$\frac{d^2 r_s}{dt^2} = -\frac{GM}{r_s^2}\left[\psi - \frac{3}{\psi c^2}\left(\frac{dr_s}{dt}\right)^2\right], \quad (5)$$

where $r_s$ is the Schwarzschild radial coordinate, $\psi(r_s) \equiv 1 - \mu/r_s$ is the $g_{00}$ component of the Schwarzschild metric, and $\mu \equiv 2GM/c^2$ is the constant Schwarzschild radius of the mass *M*. A test particle approaching or receding radially from the stationary mass *M* at a speed greater than $3^{-1/2}\psi c$ is repelled, as seen by a distant inertial observer, a result first derived by (Hilbert, 1917), as seen in Figure 1. (In Hilbert's notation, $1 - \alpha/r$ is the same as our $\psi$.)

For $\gamma \gg 1$, equation (3) shows that $g_r \approx -8\gamma^5 g_N$ in the forward ($\theta' = 0$) direction, where $g_N = -Gm/\{r^2\}_{\text{ret}}$ is the Newtonian field. For $\gamma \gg 1$, the repulsive field in the backward ($\theta' = \pi$) direction, $g_r \approx -(\gamma/2)g_N$, is much weaker. At extreme relativistic velocities, such as are attained in particle accelerators and storage rings, the repulsive gravitational field of particle bunches in their forward direction can be many orders of magnitude greater than the Newtonian field. This fact will be used in a later section to propose a test of relativistic gravity on the LHC.

## EXACT 'ANTIGRAVITY' FIELDS

An *exact* time-dependent field solution of Einstein's gravitational field equation for a spherical mass moving with arbitrarily high *constant* velocity was recently derived and analyzed (Felber, 2008, 2009). For the special case of a mass moving with constant velocity, the exact 'antigravity' field calculated in (Felber, 2008, 2009) corresponds at all source speeds to the weak 'antigravity' fields for any arbitrary velocity and acceleration derived by a retarded-potential methodology in (Felber 2005a), analyzed in (Felber, 2005b, 2006a, 2006b, 2006c), and reviewed in the preceding section. This section reviews the explicit coordinate transformations that demonstrate this correspondence.

The exact dynamic fields in (Felber, 2008) were calculated from an exact metric, which was first derived by (Hartle, Thorne and Price, 1986), but which had never before been analyzed. The exact 'antigravity-field' solutions in (Felber, 2008, 2009) confirm that any mass having a speed greater than $3^{-1/2}c$ gravitationally repels particles at rest within a forward and backward cone, no matter how light the mass or how weak its field.

The exact dynamic metric of a mass *m*, spherically symmetric in its rest frame, moving with constant velocity $\beta_0 c$ in the $+z$ direction has nonvanishing components of the metric in isotropic *t*, *x*, *y*, *z* coordinates of (Hartle, Thorne and Price, 1986)

$$g_{00} = p - \beta_0^2 q, \quad g_{03} = g_{30} = -\beta_0(p-q), \quad g_{11} = g_{22} = -q/\gamma_0^2, \quad g_{33} = \beta_0^2 p - q, \quad (6)$$

where $p \equiv \gamma_0^2[(1-\rho)/(1+\rho)]^2$, $q \equiv \gamma_0^2(1+\rho)^4$, $\rho \equiv Gm/2c^2\tilde{r}$, and $\tilde{\mathbf{r}} \equiv \tilde{\mathbf{x}} + \tilde{\mathbf{y}} + \tilde{\mathbf{z}}$ is the displacement vector from the source position to the test particle or detector at $\tilde{\mathbf{r}}(\tilde{t})$ in the 'comoving' Cartesian coordinates,

$$\tilde{t} = \gamma_0(t - \beta_0 z/c), \quad \tilde{x} = x, \quad \tilde{y} = y, \quad \tilde{z} = \gamma_0(z - \beta_0 ct). \quad (7)$$

From the exact dynamic metric in equation (6), the Christoffel symbols, $\Gamma^\alpha_{\mu\nu} \equiv (g^{\alpha\beta}/2)(\partial_\mu g_{\nu\beta} + \partial_\nu g_{\mu\beta} - \partial_\beta g_{\mu\nu})$, are calculated, where $\partial_\mu$ is the gradient operator. The gravitational field at a test particle with 4-velocity $v^\alpha$ is given by the geodesic equation, $dv^\mu/d\tau + \Gamma^\mu_{\alpha\beta}v^\alpha v^\beta = 0$, where $\tau$ is the proper time.

A simple calculation of an exact strong dynamic gravitational field from equation (6) is the field on a test particle *at rest* from a mass *m* moving along the *z* axis. Since the test particle is at rest, $v^i = 0$, where $i = 1, 2, 3$. For this special case, the spacetime interval at the test particle is $d\tau^2 = g_{00}dt^2$, so that $dt/d\tau = (g_{00})^{-1/2}$. In coordinate time *t*, therefore, the geodesic equation becomes $dv^i/dt + c^2\Gamma^i_{00} = 0$. The exact strong time-dependent gravitational field on a particle instantaneously *at rest* at $(t, x_0, y_0, z_0)$, as seen by a *distant inertial observer*, is therefore

$$\mathbf{g}(\mathbf{r_0},t) = \frac{d^2\mathbf{r_0}}{dt^2} = -\frac{\gamma_0^2 Gm}{\tilde{r}^3}\left\{\left(\frac{1-\rho}{(1+\rho)^7} + \frac{\beta_0^2}{1+\rho}\right)(\tilde{\mathbf{x}}+\tilde{\mathbf{y}}) + \left[\frac{1-\rho}{(1+\rho)^7} - \beta_0^2\left(\frac{3-\rho}{1-\rho^2}\right)\right]\gamma_0\tilde{\mathbf{z}}\right\}. \qquad (8)$$

Equation (8) is an exact time-dependent solution of Einstein's equation for the field on a particle *at rest* produced by a mass moving at constant speed along the *z* axis. In the weak-field approximation ($\rho \approx 0$), equation (8) agrees with (Felber, 2005a) and equation (2), and the 'antigravity' threshold on the *z* axis is $\beta_0 \approx 3^{-1/2}$. Equation (8) also confirms the weak-field result that the transverse component of the gravitational field is always attractive, no matter how strong the field or how fast the source. A particle at rest will only be repelled by a relativistic mass, therefore, if the particle lies within a sufficiently narrow angle to its forward or backward direction. That is, to be repelled, the particle must lie within the conical surface on which the radial component of the gravitational field vanishes.

The weak gravitational field of a mass having any arbitrary motion, equation (2), was calculated in (Felber, 2005a) using retarded coordinates. That is, the gravitational field **g** on a test particle at rest at the spacetime point $(\mathbf{r_0},t)$ was calculated from the motion of the source when it produced the field at the retarded time $t' = t - r'/c$ and at the spacetime point $(\mathbf{s'},t')$, where $\mathbf{r'} = \mathbf{r_0} - \mathbf{s'}$ is the retarded displacement vector from the source at $(\mathbf{s'},t')$ to the test particle at $(\mathbf{r_0},t)$.

For a mass moving with constant velocity $\beta_0 c$ in the *z* direction, Table 1 gives the coordinate transformations between the isotropic coordinates of the exact solution in (Felber, 2008, 2009) and the retarded (primed) coordinates of the weak-field solution in (Felber, 2005a). With these transformations, and in the weak-field approximation, equation (8) expressed in retarded coordinates becomes

$$\mathbf{g}(\mathbf{r_0},t) \approx -\left(\gamma_0 Gm/(\kappa' r')^3\right)\left[(1-\beta_0^4)(\mathbf{x'}+\mathbf{y'}) + (1-3\beta_0^2)(\mathbf{z'} - r'\boldsymbol{\beta_0})\right], \qquad (9)$$

where $\kappa' \equiv 1 - \beta_0 z'/r'$, and, as before, a prime denotes a quantity evaluated at the retarded time $t'$.

**Table 1.** Transformations from isotropic to retarded (primed) coordinates and back.

| | | | |
|---|---|---|---|
| $t = t' + r'/c$ | $t' = \gamma_0^2(t - \beta_0 z/c) - \gamma_0 \tilde{r}/c$ | $z - \beta_0 ct = z' - \beta_0 r'$ | $z' + \beta_0 ct' = z$ |
| $x = x'$ | $x' = x$ | $\tilde{r} = \gamma_0(r' - \beta_0 z') \equiv \gamma_0 \kappa' r'$ | $r' = \beta_0 \gamma_0^2(z - \beta_0 ct) + \gamma_0 \tilde{r}$ |
| $y = y'$ | $y' = y$ | $dt = \kappa' dt' = -\kappa' dz'/\beta_0 c$ | $dt'/r' = \gamma_0 dt/\tilde{r}$ |
| $z = z' + \beta_0 ct'$ | $z' = \gamma_0^2(z - \beta_0 ct) + \beta_0 \gamma_0 \tilde{r}$ | $dz = 0$ | $dz'/r' = -\beta_0 \gamma_0 c dt/\tilde{r}$ |

The weak field of a mass with arbitrary velocity, first derived and analyzed in retarded coordinates in (Felber, 2005), is identical, for the special case of constant source velocity along the *z* axis, to the weak field in equation (9). Correspondence is thereby demonstrated between the weak 'antigravity' field in retarded coordinates of a mass moving with arbitrary velocity in (Felber, 2005a, 2005b, 2006a, 2006b and 2006c) and the exact 'antigravity' field in isotropic coordinates of a mass moving with constant velocity in equation (8).

## 'ANTIGRAVITY' PROPULSION

This section summarizes earlier calculations (Felber, 2005b, 2006a) of exact payload trajectories in the strong gravitational fields of compact masses moving with constant relativistic velocities. Even the weak field of a suitable driver mass at relativistic speeds can quickly propel a heavy payload from rest to a speed significantly faster than the driver, a condition called hyperdrive. This section presents hyperdrive thresholds and maxima, calculated as functions of driver mass and velocity.

Exact unbounded orbits of a payload in the gravitational field of the much greater mass of a source moving at constant velocity have been analyzed in (Felber, 2005b, 2006a, 2006b and 2006c). At relativistic speeds, a suitable driver mass can quickly propel a heavy payload from rest nearly to the speed of light with negligible stresses on the payload. The first papers (Felber, 2005b, 2006a) analyzed payload motion only for drivers and payloads approaching each other directly, with zero impact parameter and zero angular momentum in the system. This section from (Felber, 2006b, 2006c) analyzes the exact strong-field orbital dynamics of a payload in the case of a close encounter, but not a collision, of a relativistic driver with an initially stationary payload.

The difficulties of calculating motion in the exact strong gravitational field of a relativistic mass can be bypassed by a two-step procedure, without calculating the field that produced the motion. The first step is to calculate the well-known exact trajectory of a payload in the static Schwarzschild field of a stationary, spherically symmetric mass. The second step is to perform a simple Lorentz transformation from the rest frame of the driver mass to the initial rest frame of the payload when the payload is far from the driver.

This two-step procedure for calculating the propulsion of a payload from rest allows no possibility for confusion or ambiguity of coordinates, because it focuses on the observed physical payload trajectory, rather than the field that produced it. The Schwarzschild solution is a unique spherically symmetric solution of Einstein's equation for which the metric becomes asymptotically flat as the distance from the source increases to infinity. The spacetime coordinates of the Schwarzschild solution, $r_s$, $\theta$, $\phi$, and $t$, are the coordinates used by a *distant inertial observer* to observe and measure the trajectory of a payload in a Schwarzschild field. A *distant inertial observer* is defined as an unaccelerated observer in flat spacetime far beyond any gravitational influence. *Every pair of distant inertial observers of a payload trajectory must agree on the position vs. time of the payload, differing in their accounts only by a simple Lorentz transformation.* In particular, the payload trajectory observed by a *distant inertial observer* in the rest frame of the driver mass must differ only by a simple Lorentz transformation from the same trajectory observed by a *distant inertial observer* in the initial rest frame of the payload (in which the payload is still beyond the gravitational influence of the driver).

This new means of 'antigravity' propulsion addresses the major engineering challenges for near-light-speed space travel of providing enormous propulsion energy quickly without undue stresses on a spacecraft. By conventional propulsion, acceleration of a 1-ton payload to $0.9c$ requires imparting a kinetic energy equivalent to about 30 billion tons of TNT. In the 'antigravity beam' of a speeding star or compact object, however, a payload would draw its energy for propulsion from the repulsive force of the much more massive driver. Moreover, since it would be moving along a geodesic, a payload would 'float weightlessly' in the 'antigravity beam' even as it was accelerated close to the speed of light.

The spacetime interval for the spherically symmetric Schwarzschild field of a mass $M$ is $c^2 d\tau^2 = \psi c^2 dt^2 - \psi^{-1} dr_s^2 - r_s^2 d\theta^2 - r_s^2 \sin^2\theta d\phi^2$. From this interval, the equations of motion of a payload in coordinate time $t$ are found to be (Chandrasekhar, 1983; Felber, 2006b)

$$\beta_r^2 = \psi^2 - \left(1 + L^2/c^2 r_s^2\right)\psi^3/\gamma_0^2, \quad \beta_\phi = L\psi/\gamma_0 c r_s, \tag{10}$$

where $\beta_r \equiv (dr_s/dt)/c$ and $\beta_\phi \equiv r_s(d\phi/dt)/c$ are the radial and azimuthal components of the normalized payload velocity $\boldsymbol{\beta}(r_s)$ measured by a *distant inertial observer*; $\gamma_0 c^2 = \psi c^2 dt/d\tau$ is the conserved total specific energy of the payload; and $L = r_s^2 d\phi/d\tau$ is its conserved specific angular momentum. If the payload has a speed $\beta_0 c$ far from the mass $M$, where $\psi \approx 1$ and $|dr_s/dt| \gg |r_s d\phi/dt|$, then from equation (10) we find $\gamma_0^2 = 1/(1-\beta_0^2)$.

The right-hand-side of the radial equation of motion in equation (10) is a normalized conservative potential, from which is derived the exact equation of motion in a Schwarzschild field (Felber, 2005b and 2006a),

$$\frac{d^2 r_s}{dt^2} = \frac{-GM}{r_s^2}\left(\frac{3\psi^2}{\gamma_0^2} - 2\psi\right) + \frac{L^2 \psi^2}{\gamma_0^2 r_s^3}\left(\psi - \frac{3GM}{r_s c^2}\right), \tag{11}$$

For purely radial motion of a payload ($L=0$), the gravitational force is repulsive when $\beta_0^2 > 1 - 2/3\psi$. From this inequality, we see that a Schwarzschild field is always seen by a *distant inertial observer* to repel a payload moving radially whenever $\beta_0 > 3^{-1/2}$, no matter how weak the field, and a strong field is seen to repel a payload whenever it is within $3\mu$, that is, within a (Schwarzschild radial) distance $6GM/c^2$, of a stationary compact object, no matter how small $\beta_0$ is. The payload acceleration in equation (11) measured by a *distant inertial observer* is the same whether the payload moves radially inwards or outwards, because it depends only on the square of $\beta_0$, and not on its sign.

Animated solutions accompanying (Felber, 2006b) graphically illustrate the 'antigravity' threshold for purely radial motion at $\beta_0 = 3^{-1/2}$, even in weak fields, found by the two-step approach presented in (Felber, 2006b). Figure 4 illustrates the two-step approach of (Felber, 2006b) to calculating the exact unbound orbital motion of a payload mass $m$ in the field of a driver mass $M$ with constant velocity $c\boldsymbol{\beta_0}$. First, the unique trajectory in the static Schwarzschild field of a stationary mass $M$ is found for which the periapsis of the payload is $b$ and the asymptotic velocity far from the stationary mass is $-c\boldsymbol{\beta_0}$. In the static field of $M$, the trajectory is symmetric about $b$ and time-reversible. Second, the trajectory is Lorentz-transformed to a reference frame moving with constant velocity $-c\boldsymbol{\beta_0}$, in which the mass $M$ has a constant velocity $c\boldsymbol{\beta_0}$, and the payload far from the driver is *initially* at rest. The Lorentz transformation occurs between two *distant inertial observers* far from the interaction, in asymptotically flat spacetime. In the dimensionless coordinate, $\rho_s \equiv b/r_s$, the orbit equation of the payload in the Schwarzschild field is (Chandrasekhar, 1983; Ohanian and Ruffini, 1994; Felber, 2006b).

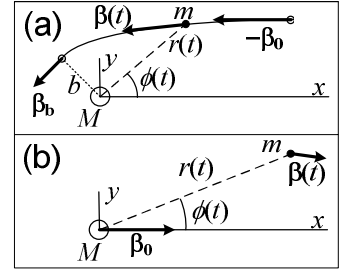

**Figure 4.** Two-step exact solution of payload trajectories: (a) Schwarzschild solution; (b) Lorentz transformation.

$$\left(\frac{d\rho_s}{d\phi}\right)^2 = (1-\sigma)\left(\frac{\gamma_0^2 - 1 + \sigma \rho_s}{\gamma_0^2 - 1 + \sigma}\right) - (1-\sigma\rho_s)\rho_s^2, \qquad (12)$$

where $\sigma \equiv \mu/b$, ranging from 0 to 1, is a measure of the field strength at periapsis, where $\rho_s = 1$. Integrating equation (12) gives the exact payload trajectories in Schwarzschild coordinates. Animated trajectory calculations, available at (Felber, 2006b), clock the payload speeds along each trajectory seen by *distant inertial observers* in both reference frames, but in Schwarzschild coordinates.

Equation (12) is the orbit equation in the Schwarzschild radial coordinate $r_s$. This orbit equation is transformed to (comoving) isotropic coordinates by the coordinate transformation (Ohanian and Ruffini, 1994), $r_s = \tilde{r}(1+\rho)^2$, where $\rho$ was defined after equation (6). The equation does not have to be transformed, however, to calculate final payload velocities. The final payload velocity is realized only when the payload is far from the driver in asymptotically flat spacetime, where $r_s = \tilde{r}$. Equation (12) may therefore be integrated, as is, to find the total deflection $\Delta\phi$ in a Schwarzschild field. The total deflection together with the final speed gives the final payload velocity.

Since trajectories are time reversible in a Schwarzschild field, the final payload speed equals the initial speed, $\beta_0$. A total deflection of 180° results in a complete U-turn and final payload velocity of $\beta_0$ in the +x direction. When Lorentz-transformed, this U-turn trajectory in a Schwarzschild field results in the maximum final hyperdrive velocity, $\beta_{\max} = 2\beta_0/(1+\beta_0^2)$, in the +x direction. This maximum hyperdrive velocity imparted to the payload corresponds to a relativistic factor, $\gamma_{\max} = (1+\beta_0^2)\gamma_0^2$, much greater than the relativistic factor of the driver mass, $\gamma_0$, if the driver speed is close to the speed of light.

The threshold field strength for hyperdrive is taken to be the value of $\sigma$ that results in propulsion of a payload from rest to the same speed, $\beta_0$, as the driver mass. Figure 5 shows the field-strength parameter $\sigma$ needed to impart to an initially stationary payload the threshold hyperdrive speed, $\beta_0$, and the maximum hyperdrive speed, $\beta_{\max} = 2\beta_0/(1+\beta_0^2)$, as a function of $\beta_0$.

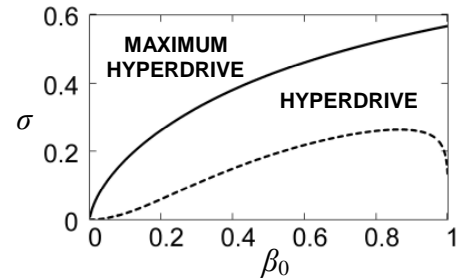

**Figure 5.** Field-strength $\sigma$ required for hyperdrive (dashed) and maximum hyperdrive (solid) vs. driver $\beta_0$.

Integrating equation (12) in the nonrelativistic Newtonian limit, in which $\beta_0^2 \ll 1$ and $\sigma \ll 1$, gives the total deflection, $\Delta\phi$, of an unbounded payload orbit in a Schwarzschild field as $\sin(\Delta\phi/2) = (1 + 2\beta_0^2/\sigma)^{-1}$. In this limit, the final speed that is payload by a moving driver is $\beta = 2\beta_0/(1+2\beta_0^2/\sigma)$. In this nonrelativistic Newtonian limit, therefore, the maximum final speed that can be imparted in the initial rest frame of the payload is $\beta_{\max} = \beta_0$, which occurs for $\beta_0 = (\sigma/2)^{1/2}$ and $\Delta\phi = 60°$.

Integrating equation (12) in the limit $\sigma \ll \beta_0^2$, which allows relativistic velocities, gives the total deflection of an unbound payload orbit in a weak Schwarzschild field as $\Delta\phi = \sigma(1 + 1/\beta_0^2)$. For $\beta_0 = 1$, this corresponds to the deflection of a photon in a weak field. In the limit of highly relativistic driver speeds ($\gamma_0 \gg 1$), even a weak field can accelerate a payload from rest to the same speed as the driver. In this limit, the threshold field strength needed for hyperdrive is only $\sigma = (2\gamma_0)^{-1/2}$. As shown in Figure 5, however, a strong field of $\sigma > 0.568$ is needed for an ultrarelativistic driver to accelerate a payload from rest to the maximum hyperdrive speed.

Although the calculation is beyond the scope of this paper, we expect that even higher payload speeds can be achieved in the strong field of a spinning compact mass. Inertial frame dragging in the strong Kerr field of a rotating compact mass (Kerr, 1963; Boyer and Lindquist, 1967; Misner, Thorne and Wheeler, 1973; Chandrasekhar, 1983; Ohanian and Ruffini, 1994) should impart angular momentum and energy to a payload orbiting in the same direction as the rotation.

## TEST OF RELATIVISTIC GRAVITY AT LHC

The strong dependence of longitudinal impulse on $\gamma$ offers opportunities for laboratory tests of gravity at relativistic velocities. Several methods should be able to provide accurate impulse measurements for the purpose of testing general relativity and discriminating among competing theories of gravity. For example, as suggested in (Braginsky, Caves and Thorne, 1977), periodic gravitational impulses delivered by proton bunches circulating in a storage ring could be measured by detectors resonant at the bunch frequency. Because the signal strength scales for $\gamma \gg 1$ as $\gamma^5$, and because the Large Hadron Collider (LHC) is much more powerful than the Tevatron-scale collider considered in (Braginsky, Caves and Thorne, 1977), the signal strengths estimated here are orders of magnitude higher, and the experiment is much more feasible now, than projected by (Braginsky, Caves and Thorne, 1977) over 30 years ago.

Comparing the phase of a stress wave in the detector to the phase of a proton bunch in the ring could give the first direct evidence of gravitational repulsion. Such an experiment to detect 'antigravity' impulses for the first time and to test relativistic gravity can be performed at the LHC off-line at any point in the tunnel, causing no interference with normal operations. For maximum effect, the resonant detector should be positioned in the plane of the proton ring at a distance *b* as close as practical to the beam axis, so that the 'antigravity beam' sweeps across the detector at the closest range.

From equation (2), one finds that the gravitational 'velocity field' of the LHC proton bunches at a detector is much greater than the 'acceleration field'. This means that during the moment that the 'antigravity beam' of each proton bunch sweeps across the detector, *the proton bunch can be treated as though it has constant velocity*. Expanding equation (2) or equation (9) for $\gamma_0 \gg 1$ and $\theta' \ll 1$ gives the weak retarded gravitational field near the forward direction of an ultrarelativistic source moving with constant velocity at a retarded distance $z'$ as

$$\mathbf{g}(\mathbf{z}', \theta') \approx \mathbf{g_0}(z')\left[(1 - \gamma_0^2 \theta'^2)/(1 + \gamma_0^2 \theta'^2)^3\right], \tag{13}$$

where $\mathbf{g_0}(z') \equiv +8\gamma_0^5 (Gm/z'^2)\hat{\mathbf{z}}$ is the field in the instantaneous forward direction.

Equation (13) was derived from the relativistically exact gravitational field in the weak-field approximation given by equation (2). The exact gravitational field solution of Einstein's equation in equation (8), however, gives an identical result. As shown in Figure 6, the exact solution in equation (8) corresponds to the weak-field solution in equation (2) for the special case of the *weak field* of an *ultrarelativistic* source moving with *constant velocity*. This correspondence was demonstrated in equation (9) by applying the coordinate transformations of Table 1 to the exact solution.

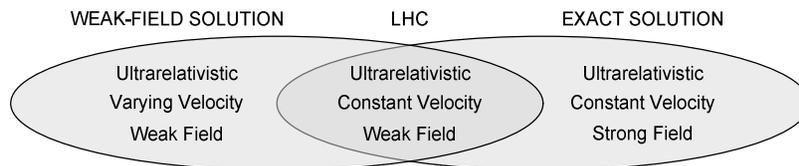

**Figure 6.** Weak-field and exact solutions predict same 'antigravity' signal strength at LHC.

From equation (13), Figure 7 shows the angular distribution (and radial profile) of the 'antigravity beam' of the ultrarelativistic proton bunches near the forward direction of the bunches. In the configuration shown in Figure 8 for a test at the LHC, the 'antigravity beam' irradiates the detector in the forward direction of the proton bunch when the bunch is a retarded distance of $z' \approx (2bR_0)^{1/2}$ from the detector, where $R_0$ is the ring radius. From equation (13), the peak repulsive field of the bunch on the detector at that distance is $g_z \approx 4\gamma_0^5 Gm/bR_0$. Since the divergence of the 'antigravity beam' is about $1/\gamma_0$, the beam sweeps across a point on the detector in a time $\Delta t \approx R_0/\gamma_0 c$, and the specific impulse delivered to the detector by a single bunch is $g_z \Delta t \approx 4\gamma_0^4 Gm/bc$. The 'antigravity beam' spot size at the detector is $z'/\gamma_0$.

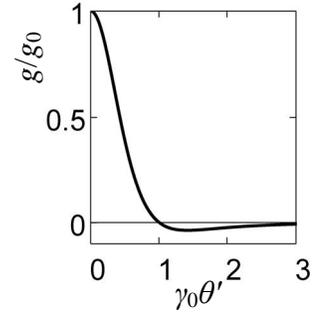

**Figure 7.** Radial profile of 'antigravity beam'.

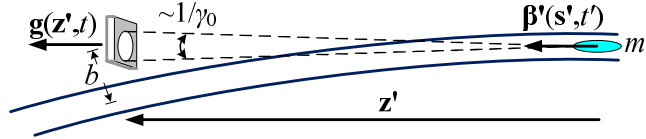

**Figure 8.** 'Antigravity beam' of proton bunch sweeping across detector (not to scale).

For $n$ equally spaced proton bunches in the ring, the bunch frequency and the impulse frequency at the detector is $f = nf_0$, where $f_0 = c/2\pi R_0$ is the circulation frequency. The duty cycle of the impulse is about $f\Delta t$, and the effective root-mean-square (rms) 'antigravity' wave amplitude is $g_{rms} \approx g_z f\Delta t$. The effective rms pressure $P_{rms}$ at the base of the detector is $g_{rms}$ times the areal mass density. For a resonant detector $N$ acoustic wavelengths thick, $P_{rms} \approx NZg_z\Delta t$, where $Z$ is the characteristic acoustic impedance of the detector. For a single proton bunch, the effective sound pressure level (SPL) of the signal is $SPL = 20\log_{10}(P_{rms}/1\ \mu Pa)$ dB $re$ 1 $\mu$Pa.

The change in velocity of the detector face produced by a single bunch is just the specific impulse, $g_z\Delta t$. The change in displacement amplitude of the face (of a resonant detector) produced by a single bunch is $\varepsilon \approx g_z\Delta t/2\pi f$. The steady-state resonant displacement amplitude is about $Q\varepsilon$, where $Q$ is the quality factor of the detector oscillator. The steady-state resonant amplitude is neared only after about $Q$ resonant oscillations of the detector.

Table 2 displays the relevant storage-ring and peak-luminosity beam parameters of the LHC (Poole, 2004). For an impact parameter $b \approx 10$ cm, each bunch irradiates the detector with an 'antigravity beam' of amplitude $g_z \approx 3$ nm/s$^2$ at a standoff range of $z' \approx 30$ m with a spot size of about $z'/\gamma_0 \approx 4$ mm. [In comparison, (Braginsky, Caves and Thorne, 1977) estimated a signal strength of order $10^{-21}$ m/s$^2$ for a Tevatron-scale collider.] The specific impulse delivered to a detector by a single bunch during exposure to the 'antigravity beam' for $\Delta t \approx 2$ ns is $5\times10^{-18}$ m/s.

**Table 2.** LHC storage-ring and peak-luminosity beam parameters (Poole, 2004).

| Parameter | Value | Parameter | Value |
|---|---|---|---|
| Ring circumference, $2\pi R_0$ (km) | 26.659 | Revolution frequency, $f_0$ (kHz) | 11.245 |
| Number of protons per bunch | 1.15x10$^{11}$ | Number of bunches, $n$ | 2808 |
| Proton energy (GeV) | 7000 | Relativistic gamma, $\gamma$ | 7461 |

When all 2808 rf buckets are filled with proton bunches, the bunch frequency and impulse frequency at the detector is $f = 31.576$ MHz. The duty cycle is $f\Delta t \approx 0.06$. A quartz crystal has a characteristic acoustic impedance $Z = 15$ MPa·s/m. The effective SPL of the 'antigravity beam' at the base of an $N$-wavelength-thick quartz detector, therefore, is $SPL \approx -80 + 20\log_{10}(N)$ dB $re$ 1 $\mu$Pa.

Near steady-state resonance, the SPL is amplified by about $20\log_{10}(Q)$ dB. Since in normal operation at LHC, the beam can circulate for 10 to 24 hours with a bunch frequency over 30 MHz, $Q$ could be as high as $10^{12}$. Such a $Q$ is well within the theoretical limits of sapphire monocrystals that (Braginsky, Caves and Thorne, 1977) suggests could be used for such an experiment. With a suitable high-$Q$ resonant detector and a typical proton circulation time of 10 hours, the SPL of the 'antigravity beam' at the LHC could be resonantly amplified to exceed 160 dB $re$ 1 $\mu$Pa.

# SUMMARY AND CONCLUSIONS

In the weak-field approximation of general relativity, the relativistically exact gravitational field of a particle having any velocity and acceleration is given in equation (2) by (Felber, 2005a). This exact weak-field solution is the first to show that a *distant inertial observer* sees masses with velocity greater than $3^{-1/2}c$ repel stationary particles within an 'antigravity beam' in the forward and backward directions. This result should perhaps not be surprising, since (Hilbert, 1917) showed over 90 years ago that a *distant inertial observer* sees a stationary mass repel particles moving radially towards it or away from it with a velocity greater than $3^{-1/2}c$. That is, it should not be surprising to anyone who believes that if A repels B, then, by conservation of momentum and mass dipole moment, B should repel A.

At high Lorentz factors ($\gamma \gg 1$), the force of repulsion in the forward direction is about $-8\gamma^5$ times the Newtonian force. The strong scaling of impulse with $\gamma$ should make certain laboratory tests of relativistic gravity much less difficult than earlier imagined by (Braginsky, Caves and Thorne, 1977). This paper outlined one such test that could be performed off-line at the LHC without interfering with any of the operations of the facility. The experiment would measure the repulsive gravitational impulses of proton bunches delivered in their forward direction to resonant detectors just outside the beam pipe. This test could provide accurate measurements of post-Newtonian parameters and the first observation of 'antigravity', as well as validating the potential utility of relativistic gravity for spacecraft propulsion in the distant future.

A new exact time-dependent field solution of Einstein's equation is given in equation (8) by (Felber, 2008 and 2009). This exact strong-field solution provides further support for the weak-field results presented in this paper. According to Table 1 and equation (9), the exact field solution in equation (8) for a mass moving with constant velocity corresponds precisely in the weak-field approximation to the weak-field solution in equation (2), for the special case of constant velocity.

A simple Lorentz transformation of the well-known unbound orbit of a payload in a Schwarzschild field gives the exact payload trajectory in the strong field of a relativistic driver with constant velocity, as seen by a *distant inertial observer*. The calculations of these payload trajectories by this two-step approach, and their animated versions (Felber, 2006b), clearly show that suitable drivers at relativistic speeds can quickly propel a heavy payload from rest to speeds close to the speed of light.

The strong field of a compact driver mass can even propel a payload from rest to speeds faster than the driver itself – a condition called hyperdrive. Hyperdrive is analogous to the elastic collision of a heavy mass with a much lighter, initially stationary mass, from which the lighter mass rebounds with about twice the speed of the heavy mass. Hyperdrive thresholds and maxima were calculated and shown in Figure 5 as functions of driver mass and velocity. Substantial payload propulsion can be achieved in weak driver fields, especially at relativistic speeds.

The exact time-dependent gravitational-field solutions of Einstein's equation in (Felber, 2008 and 2009) for a mass moving with constant velocity, and the two-step approach in (Felber, 2005b, 2006a, 2006b and 2006c) to calculating exact orbits in dynamic fields, and the retarded fields calculated in (Felber, 2005a) all give the same result: Even weak gravitational fields of moving masses are repulsive in the forward and backward directions at source speeds greater than $3^{-1/2}c$.

The field solutions in this paper have potential theoretical and experimental applications in the near term and potential propulsion applications in the long term. In the near term, the solutions can be used in the laboratory to test relativistic gravity for the first time. Performing such a test at an accelerator facility has many advantages over similar space-based tests of relativity that have been performed and contemplated for the future, including low cost, quickness, convenience, ease of data acquisition and data processing, and an ability to modify and iterate tests in real time. Such a test could provide accurate measurements of post-Newtonian parameters in the extreme relativistic regime and the first observation of 'antigravity'. Our estimates suggest that each proton bunch in the LHC beam would produce an 'antigravity beam' with a signal strength of 3 nm/s$^2$ and a duration of 2 ns at a detector. With a suitable high-$Q$ resonant detector, a typical proton circulation time of 10 hours, and an impulse frequency at peak luminosity of 31.6 MHz, the SPL of the 'antigravity beam' at the LHC could be resonantly amplified to exceed 160 dB *re* 1 µPa.

Although not discussed here, the possibility exists that the first observation of 'antigravity' could be made at the Tevatron at the Fermi National Accelerator Laboratory in Batavia, Illinois. The experiment would be more challenging at the Tevatron, owing primarily to the $\gamma^4$ dependence of the 'antigravity' impulse of the proton bunches, which by itself gives the LHC an advantage of a factor of 2400 in signal strength. Nevertheless, a window of opportunity is open while the Tevatron is fully operational, before it is decommissioned.

In the long term, gravitational repulsion at relativistic speeds opens vistas of opportunities for spacecraft propulsion (Felber, 2005b, 2006a, 2006b and 2006c). The use of 'antigravity' for spacecraft propulsion is appealing on two counts. As difficult as it is now to imagine how to access the energy of a large relativistic driver mass in deep space, it seems no less difficult to imagine how otherwise to provide the kinetic energy equivalent of 30 billion tons of TNT needed to accelerate a 1-ton payload to $0.9c$. Particularly appealing is that propulsion of a massive payload to relativistic speeds can be accomplished by a relativistic driver quickly and with manageable stresses, because the only stresses in acceleration along a geodesic arise from tidal forces much weaker than the propulsion forces.

## NOMENCLATURE

| | | | |
|---|---|---|---|
| $b$ | = impact parameter, periapsis (cm) | $\alpha$ | = $2\gamma - 1/\gamma$ (unitless) |
| $c$ | = speed of light (cm/s) | $\boldsymbol{\beta}$ | = source velocity normalized to $c$ (unitless) |
| $d\tau^2$ | = spacetime interval (s$^2$) | $\gamma$ | = Lorentz (relativistic) factor (unitless) |
| $f$ | = proton bunch frequency (Hz) | $\Gamma^\alpha_{\mu\nu}$ | = Christoffel symbol (cm$^{-1}$) |
| $f_0$ | = proton circulation frequency (Hz) | $\delta$ | = delta function (inverse unit of argument) |
| $\mathbf{g}$ | = gravitational acceleration 3-vector (cm/s$^2$) | $\varepsilon$ | = incremental displacement amplitude (cm) |
| $\mathbf{g_0}$ | = gravitational field in forward direction (cm/s$^2$) | $\Delta t$ | = time interval (s) |
| $g_N$ | = magnitude of Newtonian field (cm/s$^2$) | $\Delta\phi$ | = angular deflection (radian) |
| $g_{\mu\nu}$ | = covariant metric tensor (unitless) | $\partial_\mu$ | = gradient operator 4-vector (cm$^{-1}$) |
| $G$ | = gravitational constant (cm$^3$/s$^2$g) | $\eta_{\mu\nu}$ | = Minkowski metric tensor (unitless) |
| $h_{\mu\nu}$ | = linearized part of metric tensor (unitless) | $\theta$ | = polar angle (radians) |
| $L$ | = specific angular momentum (cm$^2$/s) | $\kappa$ | = directional Doppler-like factor (unitless) |
| $m$ | = source rest mass (g) | $\mu$ | = constant Schwarzschild radius (cm) |
| $M$ | = large stationary-source rest mass (g) | $\rho$ | = normalized comoving inverse radius (unitless) |
| $\hat{\mathbf{n}}$ | = unit vector in direction of $\mathbf{r}$ (unitless) | $\rho_s$ | = normalized Schwarzschild inverse radius (unitless) |
| $n$ | = number of proton bunches (unitless) | $\sigma$ | = gravitational field strength parameter (unitless) |
| $N$ | = number of acoustic wavelengths (unitless) | $\tau$ | = proper time (s) |
| $p$ | = quantity in exact metric (unitless) | $\phi$ | = azimuthal coordinate (radian) |
| $P$ | = pressure (dyn/cm$^2$) | $\psi$ | = $g_{00}$ component of Schwarzschild metric (unitless) |
| $q$ | = quantity in exact metric (unitless) | | **Notation Conventions** |
| $\mathbf{r}$ | = displacement vector – source to detector (cm) | | Boldface denotes 3-vector |
| $\mathbf{r_0}$ | = displacement vector – origin to detector (cm) | | Prime denotes retarded quantity |
| $r_s$ | = Schwarzschild radial coordinate (cm) | | $\{\ \}_{\text{ret}}$ denotes retarded quantity inside brackets |
| $R_0$ | = storage-ring radius (cm) | | Twiddle denotes comoving coordinate |
| $\mathbf{s}$ | = displacement vector – origin to source (cm) | | Overdot denotes differentiation with respect to $t'$ |
| $S_{\mu\nu}$ | = source tensor (erg/g) | | Cap denotes unit vector |
| $SPL$ | = sound pressure level (dB $re$ 1 µPa) | | Subscript 0 denotes constant quantity |
| $t$ | = time (s) | | Subscript $b$ denotes periapsis |
| $\mathbf{u}$ | = velocity 3-vector of source (cm/s) | | Subscript $c$ denotes critical threshold |
| $u^\mu$ | = velocity 4-vector of source (cm/s) | | Subscript max denotes maximum quantity |
| $\mathbf{v}$ | = velocity 3-vector of test particle (cm/s) | | Subscript $r$ denotes radial component |
| $v^\mu$ | = velocity 4-vector of test particle (cm/s) | | Subscript rms denotes effective root-mean-square quantity |
| $x,y,z$ | = isotropic Cartesian coordinates (cm) | | Subscripts $x,y,z$ denote $x,y,z$ components |
| $Z$ | = characteristic acoustic impedance (g/cm$^2$s) | | Subscript $\phi$ denotes azimuthal component |

## ACRONYMS

LHC – Large Hadron Collider
NASA – National Aeronautics and Space Administration
SPL – sound pressure level

# ACKNOWLEDGMENTS

I am grateful to Julian Moore and Anastasios Kartsaklis for bringing to my attention early references to Hilbert's calculation of gravitational repulsion in a Schwarzschild field.